\documentclass[journal]{IEEEtran}
\IEEEoverridecommandlockouts

\usepackage{cite}
\usepackage{amsmath,amssymb,amsfonts}
\usepackage{mathtools}
\usepackage{amsthm}
\usepackage{algorithm}
\usepackage{algpseudocode}
\usepackage{etoolbox}
\usepackage{graphicx,color}
\usepackage{textcomp}
\usepackage[dvipsnames,rgb]{xcolor}
\usepackage{makecell}
\usepackage{colortbl}
\usepackage{float}
\usepackage{tikz}
\usetikzlibrary{arrows.meta, positioning, shapes.geometric,decorations.pathreplacing}
\usetikzlibrary{patterns}
\usepackage{siunitx}
\usepackage{tabularx}
\usepackage{multirow}
\usepackage{array}
\usepackage{acronym}
\usepackage{subcaption}
\usepackage{booktabs}
\usepackage{hyperref} 
\usepackage[capitalize,noabbrev]{cleveref}
\usepackage[textsize=tiny]{todonotes}
\usepackage{latexsym}
\usepackage{gensymb}
\usepackage{enumitem}
\usepackage{esvect}
\usepackage{comment}
\usepackage[normalem]{ulem}
\usepackage{soul}
\usepackage{kotex}
\usepackage{romannum}
\usepackage{caption}
\usepackage{breqn}
\usepackage{rotating}
\usepackage{pdflscape}

\input{defs/macros}
\theoremstyle{plain}

\theoremstyle{definition}

\theoremstyle{remark}

\definecolor{nyuviolet}{RGB}{87, 6, 140}

\def\BibTeX{{\rm B\kern-.05em{\sc i\kern-.025em b}\kern-.08em
    T\kern-.1667em\lower.7ex\hbox{E}\kern-.125emX}}

\newacro{ACDD}{Alamouti with cyclic delay diversity}
\newacro{URLLC}{ultra-reliable low-latency communications}
\newacro{3GPP}{third generation partnership project}
\newacro{PHY}{physical layer}
\newacro{MIMO}{multiple-input multiple-output}
\newacro{MU-MIMO}{multi-user multiple-input multiple-output}
\newacro{SIMO}{single-input multiple-output}
\newacro{MISO}{multiple-input single-output}
\newacro{SISO}{single-input single-output}
\newacro{MRC}{maximum-ratio combining}
\newacro{SNR}{signal-to-noise ratio}
\newacro{CP}{cyclic prefix}
\newacro{CDD}{cyclic delay diversity}
\newacro{FSC}{frequency-selective channel}
\newacro{STC}{space-time coding}
\newacro{FFT}{fast Fourier transform}
\newacro{LMMSE}{linear minimum mean-squared error}
\newacro{CFAC}{cross frequency AoA consistency}
\newacro{FER}{frame error rate}
\newacro{OFDM}{orthogonal frequency division multiplexing}
\newacro{OCDM}{orthogonal chirp division multiplexing}
\newacro{RMS}{root mean square}
\newacro{DS}{delay spread}
\newacro{FSC}{frequency-selective channel}
\newacro{CSI}{channel state information}
\newacro{LMMSE-PIC}{linear minimum mean squared error with parallel interference cancellation}
\newacro{PFE}{perfect-feedback equalizer}
\newacro{FD}{full-duplex}
\newacro{PDP}{power delay profile}
\newacro{PDF}{probability density function}
\newacro{DFT}{discrete Fourier transform}
\newacro{SDFT}{sparse DFT}
\newacro{ICI}{inter-carrier interference}
\newacro{OTFS}{orthogonal time frequency space}
\newacro{AWGN}{additive white Gaussian noise}
\newacro{SWH}{sparse Walsh-Hadamard}
\newacro{LLR}{log-likelihood ratio}
\newacro{PMF}{probability mass function}
\newacro{CRC}{cyclic redundancy check}
\newacro{PAM}{pulse amplitude modulation}
\newacro{QAM}{quadrature amplitude modulation}
\newacro{FWHT}{fast Walsh-Hadamard transform}
\newacro{MAP}{maximum a-posteriori}
\newacro{SC}{specular component}
\newacro{CFO}{carrier frequency offset}
\newacro{ISI}{inter-symbol interference}
\newacro{ZP}{zero-padding}
\newacro{EVD}{eigenvalue decomposition}
\newacro{BCJR}{Bahl, Cocke, Jelinek, and Raviv}
\newacro{WHT}{Walsh-Hadamard transform}
\newacro{APP}{a-posteriori probability}
\newacro{SILE-EPIC}{self-iterated linear equalizer with expectation propagation}
\newacro{EP}{expectation propagation}
\newacro{i.i.d.}{independent and identically distributed}
\newacro{CWCU}{component wise conditionally unbiased}
\newacro{MSE}{mean squared error}
\newacro{EXIT}{extrinsic information transfer}
\newacro{MI}{mutual information}
\newacro{PAPR}{peak-to-average power ratio}
\newacro{DFT-s}{discrete Fourier transform-spread}
\newacro{AMP}{approximate message passing}
\newacro{GAMP}{generalized \ac{AMP}}
\newacro{VAMP}{vector \ac{AMP}}
\newacro{RSC}{recursive systematic convolutional}
\newacro{QPSK}{quadrature phase-shift keying}
\newacro{CFAR}{constant false alarm rate}
\newacro{PD}{probability of detection}
\newacro{PFA}{probability of false alarm}
\newacro{RV}{random variable}
\newacro{CDF}{cumulative distribution function}
\newacro{HD-ZP}{half-duplex ZP}
\newacro{FD-CP}{full-duplex ZP}
\newacro{DFRC}{dual-function radar communication}
\newacro{SINR}{signal-to-interference noise ratio}
\newacro{ISAC}{integrated sensing and communication}
\newacro{SI}{self-interference}
\newacro{RSI}{residual self-interference}
\newacro{ADC}{analog-to-digital converter}
\newacro{DAC}{digital-to-analog converter}
\newacro{ED}{energy-detection}
\newacro{IDFT}{inverse discrete Fourier Transform}
\newacro{SFFT}{symplectic finite Fourier transform }
\newacro{CRB}{Cram{\'{e}}r-Rao bound}
\newacro{ZC}{Zadoff-Chu}
\newacro{RMSE}{root mean square error}
\newacro{MMSE}{minimum mean-square error}
\newacro{UW}{unique word}
\newacro{GFDM}{generalized frequency division multiplexing}
\newacro{RRC}{root-raised cosine}
\newacro{UB}{upper bound}
\newacro{CEF}{channel estimation field}
\newacro{TRX}{transceiver}
\newacro{IF}{intermediate frequency}
\newacro{RF}{radio frequency}
\newacro{FPGA}{field programmable gate arrays}
\newacro{SDR}{software-defined radio}
\newacro{UWB}{ultra wideband}
\newacro{FR3}{frequency range 3}
\newacro{PCB}{printed circuit board}
\newacro{SMA}{SubMiniature version A}
\newacro{MUSIC}{multiple signal classification}
\newacro{CIR}{channel impulse response}
\newacro{FR}{Frequency Range}
\newacro{mmWave}{millimeter wave}
\newacro{LoS}{line-of-sight}
\newacro{AoD}{angle-of-departure}
\newacro{ESNR}{estimation SNR}
\newacro{AoA}{angle-of-arrival}
\newacro{SDNR}{signal-to-DMC-noise ratio}
\newacro{ULA}{uniform linear array}
\newacro{DMC}{dense multipath component}
\newacro{ML}{maximum-likelihood}
\newacro{IFFT}{inverse fast Fourier transform}
\newacro{LM}{Levenberg-Marquardt}
\newacro{ACF}{autocorrelation function}
\newacro{UWB}{ultra-wideband}
\newacro{SLAM}{simultaneous localization and mapping}
\newacro{STO}{sampling time offset}
\newacro{GLRT}{generalized likelihood ratio test}
\newacro{FD-ED}{frequency-domain energy detectors}
\newacro{IOU}{intersection over union}
\newacro{FLOP}{floating point operation}
\newacro{FLOPS}{floating point operations per second}
\newacro{LTBF}{long-term beamforming}
\newacro{UE}{user equipment}
\newacro{SRS}{sounding reference signal}
\newacro{BW}{bandwidth}
\newacro{RB}{resource block}
\newacro{RE}{resource element}
\newacro{BS}{base station}
\newacro{CG}{conjugate-gradient}
\newacro{SRAM}{static random access memory}
\newacro{DSP}{digital signal processor}
\newacro{VLSI}{very large-scale integration}
\newacro{NR}{New Radio}
\newacro{ASIC}{application-specific integrated circuit}
\newacro{gNB}{next-generation NodeB}
\newacro{SCS}{subcarrier spacing}
\newacro{WMMSE}{weighted minimum mean square error}
\newacro{WSR}{weighted sum-rate}
\newacro{ZF}{zero-forcing}
\newacro{UMB}{upper mid-band}
\newacro{DM-RS}{demodulation reference signal}

\begin{document}

\title{Efficient Upper Mid-Band Spectrum Sensing with Multiple Signals}

%\author{Yongjun~Kim,~\IEEEmembership{Graduate Student Member,~IEEE},
%        Ali~Rasteh,~\IEEEmembership{Graduate Student Member,~IEEE},
%        Sundeep~Rangan,~\IEEEmembership{Fellow,~IEEE},
%        and~Junil~Choi,~\IEEEmembership{Senior Member, IEEE}% <-this % stops a space
\author{Yongjun~Kim,
        Ali~Rasteh,
        Sundeep~Rangan,
        and~Junil~Choi% <-this % stops a space
\thanks{
Y. Kim and J. Choi are with the School of Electrical Engineering, KAIST, Daejeon 34141, Korea (email: {yongjunkim, junil}@kaist.ac.kr).}
\thanks{
A. Rasteh and S. Rangan are with New York University, Brooklyn, NY (email: {ar7655, srangan}@nyu.edu).
}}

\maketitle

\thispagestyle{empty}
\pagestyle{empty}

\acresetall
\begin{abstract}
Spectrum sensing is a fundamental problem in the upper mid-band, where spectrum resources are shared with incumbent systems. This paper considers frequency-domain occupancy estimation when the number of primary user signals, their bandwidths, and their signal-to-noise ratios are all unknown. We develop a generalized likelihood ratio test and a computationally efficient search procedure that combines binary search with dynamic programming to select the set of intervals maximizing the sum of log-likelihoods. The proposed method is validated through both simulation data and over-the-air experimental data using an upper mid-band software-defined radio (SDR), demonstrating its practical applicability.
\end{abstract}

\begin{IEEEkeywords}
Spectrum sensing, upper mid-band, hypothesis testing, efficient binary search
\end{IEEEkeywords}

\IEEEpeerreviewmaketitle

\section{Introduction}
\label{Sec:Introduction}
Spectrum sensing (SS) enables a secondary user to avoid interference to a primary user (PU) by detecting unused portions of a frequency band. SS has recently attracted more attention in the upper mid-band (6–24 GHz), which has emerged as a promising candidate for next-generation wireless systems due to its balance between coverage and bandwidth availability. In this band, spectrum access often depends on coexistence with incumbents such as satellite and radar systems \cite{bazzi:2025}.

Although SS has long been a core technology of cognitive radio, many existing advanced methods rely on prior knowledge of the PU or incur substantial computational costs, which are particularly problematic for real-time operation in the upper mid-band. For instance, the optimality of the matched filter is ensured only when prior knowledge of the PU network is given \cite{ma:2012,zhang:2014}, and eigenvalue-based detectors achieve strong performance at low SNR, yet they exploit specific signal properties such as covariance structure and suffer from high computational complexity \cite{zeng:2009,kortun:2013}. While energy detection is attractive due to its simplicity, it still assumes a fixed sensing bandwidth or known band structures 
%\cite{sobron:2015,chatziantoniou:2015}. 
\cite{quan:2008}.
The work in \cite{rasteh:2026} proposed a near-optimal, low-complexity SS algorithm based on a Generalized Likelihood Ratio Test (GLRT) for non-coherent power measurements. While demonstrating high efficiency, i.e., reducing search complexity from $\mathcal{O}(N^2)$ to $\mathcal{O}(N)$, its key limitation is the assumption that the number of signals is either zero or one. In this paper, we extend \cite{rasteh:2026} to address an arbitrary number of signals while preserving near-optimality and low complexity. We retain its foundational system model---including the assumptions of known, pre-calibrated noise power and constant signal-to-noise ratio (SNR) over each occupied interval---but generalize the detection framework. Taking practical deployments into account, we assume no prior knowledge of the number of signals, their bandwidths, or their SNRs.

%\begin{figure}[t]
%\centering
%\includegraphics[width=0.7\columnwidth]{figs/system_model.png}
%\caption{Simulated power measurement per frequency bin.}
%\label{fig:system_model}
%\end{figure}

The main contributions of this work are summarized as follows:
\begin{enumerate}
\item We develop a computationally efficient binary-search-based procedure to estimate signal intervals under a practical setting where the number of signals, their bandwidths, and their SNRs are unknown. We connect SS as a weighted interval scheduling (WIS) problem and employ dynamic programming (DP) to select the interval set that maximizes the sum of log-likelihoods, together with low-complexity binary search.
\item We validate the proposed algorithm for both simulation data and real upper mid-band measurements. Specifically, we use the Pi-Radio upper mid-band software-defined radio (SDR) platform \cite{mezzavilla:2024} to demonstrate the estimation accuracy of the proposed approach in over-the-air measurements.
\end{enumerate}

The remainder of the paper is organized as follows. Sec.~\ref{Sec:Problem formulation} formulates the interval estimation problem based on a generalized likelihood ratio test (GLRT). Sec.~\ref{sec:method} presents the three-stage binary search method, including dyadic interval estimation, DP-based dyadic interval selection, and interval refinement. Sec.~\ref{sec:measurement} describes the upper mid-band measurement setup using the Pi-Radio SDR. Sec.~\ref{Sec:Results} evaluates the proposed algorithm through both simulation and measurement data. Finally, Sec.~\ref{Sec:Conclusion} concludes the paper.

\section{Problem formulation and GLRT}
\label{Sec:Problem formulation}
Consider a one-dimensional line search problem over $N$ frequency bins indexed by $n\in\{0,1,\ldots,N-1\}$, with a power measurement vector $\mathbf{X}=[X_0,X_1,\ldots,X_{N-1}]^{\mathrm{T}}$. There are $K$ signals occupying unknown, disjoint intervals $\mathcal{S}=\bigcup_{k=1}^{K}\mathcal{S}_k$, where $\mathcal{S}_k=[a_k,b_k)$ for signal index $k\in\{1,\ldots,K\}$ and $0\le a_1 < b_1 < \cdots < a_K < b_K \le N$. We assume a guard band of size $G>0$ between adjacent signals, i.e., $b_k+G\le a_{k+1}$ for $k=1,\ldots,K-1$. Following the model in \cite{rasteh:2026}, it is assumed that non-coherent power measurements $\{X_n\}_{n=0}^{N-1}$ are normalized to a unit noise variance, which requires an accurate noise estimation to be performed during a reliable calibration phase. Such an exponential model naturally arises in non-coherent detection without synchronization, where unsynchronized complex symbols and noise can be globally modeled as Gaussian processes. Consequently, the measurements are modeled as independent exponential random variables with means
\begin{align}
\mathbb{E}[X_n]=
\begin{cases}
1+\gamma_k, & n\in \mathcal{S}_k,\\
1, & n \notin \mathcal{S},
\end{cases}
\label{eq:expectation cases}
\end{align}
where $\gamma_k$ denotes the SNR of signal $k$. Furthermore, assigning a single parameter $\gamma_k$ to the interval $\mathcal{S}_k$ implicitly assumes the channel exhibits flat fading across each transmission band, yielding a constant piecewise power spectral density.
%An example of this model is shown in Fig.~\ref{fig:system_model}.

Our objective is to estimate the occupied set $\mathcal{S}$ through a GLRT,
\begin{align}
\mathcal{S}^* = \arg\max_{\mathcal{S}'}
\log \frac{p(\mathbf{X}\mid \mathcal{H}_A(\mathcal{S}') = 1)}{p(\mathbf{X}\mid \mathcal{H}_0(\mathcal{S}') = 1)},
\label{eq:p1}
\end{align}
where $\mathcal{H}_0(\mathcal{S}')$ denotes the null hypothesis that there is only noise in $\mathcal{S}'$, and $\mathcal{H}_A(\mathcal{S}')$ denotes the alternative hypothesis that signals are present on $\mathcal{S}'$. For convenience, define the hypothesis for a single arbitrary interval $\mathcal{S}_{k'} \subseteq \mathcal{S}'$ as 
\begin{align}
H_{k'}=
\begin{cases}
1, & \text{a signal is present on } \mathcal{S}_{k'},\\
0, & \text{no signal is present on } \mathcal{S}_{k'},
\end{cases}
\end{align}
so that $\mathcal{H}_A(\mathcal{S}')$ corresponds to $\{H_{k'}=1\}_{k'=1}^{K'}$, where $K'$ is the number of chosen intervals, and $\mathcal{H}_0(\mathcal{S}')$ corresponds to $\{H_{k'}=0\}_{k'=1}^{K'}$. Let $\mathbf{X}_{\mathcal{S}_{k'}} = \{X_n \mid n\in \mathcal{S}_{k'}\}$ be a sequence of power measurements over $\mathcal{S}_{k'}$. Under the independence assumption across bins, the likelihood factorizes as
\begin{align}
p(\mathbf{X}\mid \mathcal{H}_A(\mathcal{S}'))
&= \prod_{k'=1}^{K'} p(\mathbf{X}_{\mathcal{S}_{k'}}\mid H_{k'}=1),\\
p(\mathbf{X}\mid \mathcal{H}_0(\mathcal{S}'))
&= \prod_{k'=1}^{K'} p(\mathbf{X}_{\mathcal{S}_{k'}}\mid H_{k'}=0).
\end{align}
Substituting into \eqref{eq:p1} yields
\begin{align}
\mathcal{S}^* = \arg\max_{\mathcal{S}'} \sum_{k'=1}^{K'}
\log \frac{p(\mathbf{X}_{\mathcal{S}_{k'}}\mid H_{k'}=1)}{p(\mathbf{X}_{\mathcal{S}_{k'}}\mid H_{k'}=0)},
\end{align}
which leads to a maximum-log-likelihood-summation-based multiple-interval estimation problem.

The log-likelihood ratio (LLR) for interval $\mathcal{S}_{k'}$ with SNR $\gamma_{k'}$ is defined as
\begin{align}
J(\mathcal{S}_{k'},\gamma_{k'})
= \log \frac{p(\mathbf{X}_{\mathcal{S}_{k'}}\mid H_{k'}=1)}{p(\mathbf{X}_{\mathcal{S}_{k'}}\mid H_{k'}=0)} .
\end{align}
Following \cite{rasteh:2026}, maximizing the LLR over $\gamma_{k'}$ yields $J(\mathcal{S}_{k'}) = \max_{\gamma_{k'}\ge 0} J(\mathcal{S}_{k'},\gamma_{k'})$
\begin{align}
J(\mathcal{S}_{k'})
= |\mathcal{S}_{k'}|\!\left(\bar{X}_{\mathcal{S}_{k'}}^+ - 1 - \log \bar{X}_{\mathcal{S}_{k'}}^+\right),
\label{eq:log-likelihood}
\end{align}
where $|\cdot|$ denotes the size of the set, $\bar{X}_{\mathcal{S}_{k'}}^+ = \max\{\bar{X}_{\mathcal{S}_{k'}},\,1\}$ and $\bar{X}_{\mathcal{S}_{k'}} = \sum_{n\in\mathcal{S}_{k'}} X_n / |\mathcal{S}_{k'}|$, which is achieved when $\gamma_{k'} = \max \{\bar{X}_{\mathcal{S}_{k'}}-1,\,0\}$. The resulting GLRT decision rule is now only dependent on the average power over the interval $\mathcal{S}_{k'}$ as
\begin{align}
\hat{H}_{k'}=
\begin{cases}
1, & \bar{X}_{\mathcal{S}_{k'}}^+ \ge u_{|\mathcal{S}_{k'}|},\\
0, & \bar{X}_{\mathcal{S}_{k'}}^+ < u_{|\mathcal{S}_{k'}|},
\end{cases}
\label{eq:local glrt}
\end{align}
where the threshold $u_{|\mathcal{S}_{k'}|}$ is bounded to strictly satisfy a target false-alarm probability constraint. As rigorously derived in \cite{rasteh:2026}, applying a union bound over the $\approx N^2/2$ possible contiguous intervals bounds the false alarms by modeling the normalized energy over pure noise bins as unscaled chi-squared distributions, leading to the required theoretical threshold:
\begin{align}
u_{|\mathcal{S}_{k'}|}
= \frac{1}{2|\mathcal{S}_{k'}|}\,
F^{-1}\!\left(\frac{2P_{\mathrm{FA}}}{N^2};\,2|\mathcal{S}_{k'}|\right).
\label{eq:threshold}
\end{align}
Here, $P_{\mathrm{FA}}$ is the target overall false-alarm probability and $F^{-1}(s;\nu)$ denotes the inverse complementary cumulative distribution function of a chi-square random variable with $\nu$ degrees of freedom.

\section{Dynamic Programming-based Efficient Spectrum Sensing}
\label{sec:method}
\subsection{Dynamic Programming}
\label{subsec:DP}
Let $\tilde{\mathcal{S}}$ denote a finite set of candidate intervals. To reduce complexity, we initially restrict the elements of $\tilde{\mathcal{S}}$ to be dyadic intervals of the form $I_{m,j} = [j\cdot2^{m},(j+1)\cdot2^{m}) \subset [0,N)$, where $m,j\in\mathbb{Z}$. An interval $I_{m,j}$ is included in $\tilde{\mathcal{S}}$ only if it satisfies $\bar{X}_{I_{m,j}}^+ \ge u_{|I_{m,j}|}$ as derived in Sec.~\ref{Sec:Problem formulation}. Given $\tilde{\mathcal{S}}$, our goal is to select a subset of non-overlapping intervals that maximizes the sum of log-likelihood gain
\begin{align}
\mathcal{D}^* = \arg\max_{\mathcal{D}\subseteq \tilde{\mathcal{S}}}
\sum_{I_{m,j} \in \mathcal{D}} J(I_{m,j})
\label{eq:DP}
\end{align}
with an additional guard-band constraint of length $G>0$ between any two selected intervals. Problem \eqref{eq:DP} is a WIS problem, which can be solved efficiently via DP. Simply index the candidates as $I_i=[\tilde{a}_i,\tilde{b}_i)$ for $i\in\{1,\ldots,|\tilde{\mathcal{S}}|\}$, sorted in non-decreasing order of the right boundary $\tilde{b}_1 \le \tilde{b}_2 \le \cdots \le \tilde{b}_{|\tilde{\mathcal{S}}|}$.
Define
\begin{align}
p(i) = \max\{\ell<i:\ \tilde{b}_\ell + G \le \tilde{a}_i\},
\end{align}
as the index of the last interval that is not overlapping with $I_i$ under the guard-band constraint. Let $V(i)$ denote the maximum achievable objective value using only the first $i$ intervals. By the principle of optimality, $V(i)$ satisfies the WIS recursion
\begin{align}
V(i) = \max\Big\{V(i-1),\ J(I_i) + V(p(i))\Big\},
\quad i=1,\ldots,|\tilde{\mathcal{S}}|,
\label{eq:WIS_recursion}
\end{align}
with initialization $V(0)=0$. If $J(I_i) + V(p(i))$ is chosen, the corresponding index $i$ is memorized to be accepted by setting $\mathcal{A}(i) = 1$, otherwise, $\mathcal{A}(i) = 0$, where $\mathcal{A}$ is initialized to be $\emptyset$. The optimally selected set $\mathcal{D}^*$ is then recovered by backtracking from $i=|\tilde{\mathcal{S}}|$. If $\mathcal{A}(i) = 1$, add $I_i$ to $\mathcal{D}^*$ and resume the backtracking at $p(i)$; otherwise, resume with $i-1$.

\begin{algorithm*}[ht]
\caption{DP-based Multi-Signal Interval Estimation}
\label{Alg:estimation}
\begin{minipage}[t]{0.48\textwidth}
\begin{algorithmic}[1]
\State \textbf{Input:} Power measurements $\mathbf{X}$, false-alarm probability $P_{\mathrm{FA}}$, guard length $G$, number of frequency bins $N$
\State \textbf{Output:} Estimated occupied intervals $\hat{\mathcal{S}}$
\State Initialize $\tilde{\mathcal{S}}, \hat{\mathcal{S}}, \mathcal{J},\mathcal{U}, \mathcal{D}^*, \mathcal{A} \gets \emptyset$, $V(0)\gets0$
\State \textit{\textbf{Step 1}: Construct dyadic candidate set $\tilde{\mathcal{S}}$ via GLRT}
\For{$\ell = 2^0 ,2^1,2^2,\ldots,2^{\log_2N}$}
    \If{$u_{\ell}\notin \mathcal{U}$}
        \State Compute $u_{\ell}$ using \eqref{eq:threshold}, $\mathcal{U}\gets \mathcal{U}\cup\{u_{\ell}\}$
    \EndIf
    \For{$a=0,\ell,2\ell,\ldots,N-\ell$}
        \State Compute $\bar{X}_{[a,b)} = \sum_{n=a}^{b-1}X_n / \ell$, where $b= a+\ell$
        \If{$\bar{X}_{[a,b)} \ge u_{\ell}$}
            \State Compute $J([a,b))$ using \eqref{eq:log-likelihood}
            \State $\tilde{\mathcal{S}}\gets \tilde{\mathcal{S}}\cup\{[a,b)\}$, $\mathcal{J}\gets \mathcal{J}\cup\{J([a,b))\}$
        \EndIf
    \EndFor
\EndFor
\State \textit{\textbf{Step 2}: Select non-overlapping intervals by DP}
\State Sort $\tilde{\mathcal{S}}=\{I_i=[\tilde{a}_i,\tilde{b}_i)\}_{i=1}^{|\tilde{\mathcal{S}}|}$ as $\tilde{b}_1\le \tilde{b}_2\le \cdots \le \tilde{b}_{|\tilde{\mathcal{S}}|}$
\State Compute $p(i)=\max\{j<i:\ \tilde{b}_j+G\le \tilde{a}_i\}$ for all $i$
\For{$i=1,2,\ldots,|\tilde{\mathcal{S}}|$}
    \If{$V(i-1) \ge J(I_i)+V(p(i))$}
        \State $V(i)\gets V(i-1)$, $\mathcal{A}(i) = 0$
    \Else
        \State $V(i)\gets J(I_i)+V(p(i))$, $\mathcal{A}(i) = 1$
    \EndIf
\EndFor
\algstore{MSIE}
\end{algorithmic}
\end{minipage}\hfill
\begin{minipage}[t]{0.48\textwidth}
\begin{algorithmic}[1]
\algrestore{MSIE}
\State Set $i\gets |\tilde{\mathcal{S}}|$
\While{$i\ge 1$}
    \If{$\mathcal{A}(i) = 1$}
        \State $\mathcal{D}^*\gets \mathcal{D}^*\cup\{I_i\}$, $i\gets p(i)$
    \Else
        \State $i\gets i-1$
    \EndIf
\EndWhile
\State \textit{\textbf{Step 3}: Refine interval boundaries}
\For{each $I_{i'}=[\tilde{a}_{i'},\tilde{b}_{i'})\in \mathcal{D}^*$}
    \State $a^*\gets \tilde{a}_i$, $b^*\gets \tilde{b}_i$, $J^*\gets J([a^*,b^*))$, $\delta\gets \tilde{b}_{i'}-\tilde{a}_{i'}$
    \While{$\delta \ge 1$}
        \For{each $a\in\{a^*-\delta,\ a^*,\ a^*+\delta\}$}
            \If{$0\le a < b^*$ \textbf{and} $J([a,b^*))>J^*$}
                \State $a^*\gets a$, $J^*\gets J([a,b^*))$
            \EndIf
        \EndFor
        \For{each $b\in\{b^*-\delta,\ b^*,\ b^*+\delta\}$}
            \If{$a^* < b \le N$ \textbf{and} $J([a^*,b))>J^*$}
                \State $b^*\gets b$, $J^*\gets J([a^*,b))$
            \EndIf
        \EndFor
        \State $\delta\gets \delta/2$
    \EndWhile
    \State $\hat{\mathcal{S}}\gets \hat{\mathcal{S}}\cup\{[a^*,b^*)\}$
\EndFor
\end{algorithmic}
\end{minipage}
\end{algorithm*}

\subsection{End-to-end algorithm}
The proposed end-to-end procedure consists of three stages. First, we construct the candidate dyadic interval set $\tilde{\mathcal{S}}$ by applying the GLRT across dyadic intervals, while recording the corresponding threshold values obtained through \eqref{eq:threshold} in a set $\mathcal{U}$ to prevent duplicated threshold calculation and the maximized log-likelihood ratios obtained through \eqref{eq:log-likelihood} in a set $\mathcal{J}$. Second, from $\tilde{\mathcal{S}}$ we extract a subset of non-overlapping intervals by solving the WIS problem via DP, as described in Sec.~\ref{subsec:DP}. In this stage, we first compute the DP table $\{V(i)\}_{i=1}^{|\tilde{\mathcal{S}}|}$ while recording the acceptance in $\{\mathcal{A}(i)\}_{i=1}^{|\tilde{\mathcal{S}}|}$, then obtain $\mathcal{D}^* \subseteq \tilde{\mathcal{S}}$ by backtracking $\{\mathcal{A}(i)\}_{i=1}^{|\tilde{\mathcal{S}}|}$ from $i = |\tilde{\mathcal{S}}|$, satisfying $|\mathcal{D}^*| = K'$. Finally, we refine the left and right boundaries of each interval in $\mathcal{D}^*$ by adjusting the endpoints iteratively with a sub-interval length $\delta$ in a direction that increases the log-likelihood. The full algorithm is summarized in Algorithm~\ref{Alg:estimation}. Detailed descriptions of Steps 1 and 3 are provided in \cite{rasteh:2026}.

The computational complexity of the first stage is $\mathcal{O}(N)$. In the second stage, sorting the candidate intervals and acquiring $\{p(i)\}_{i=1}^{|\tilde{\mathcal{S}}|}$ requires $\mathcal{O}(|\tilde{\mathcal{S}}|\log_2|\tilde{\mathcal{S}}|)$, and the DP recursion requires $\mathcal{O}(|\tilde{\mathcal{S}}|)$, yielding an overall complexity of $\mathcal{O}(|\tilde{\mathcal{S}}|\log_2|\tilde{\mathcal{S}}|)$ for the DP stage. Since $|\tilde{\mathcal{S}}| < 2N$, the worst-case complexity of the first two stages is $\mathcal{O}(N\log_2 N)$. The third stage has complexity $\mathcal{O}(\log_2 N)$, and therefore does not change the overall scaling. Consequently, the total complexity is bounded by $\mathcal{O}(N\log_2 N)$, which is substantially lower than the $\mathcal{O}(N^2)$ complexity of exhaustive maximum-likelihood search.

\begin{figure}[t]
\centering
\includegraphics[width=0.8\columnwidth]{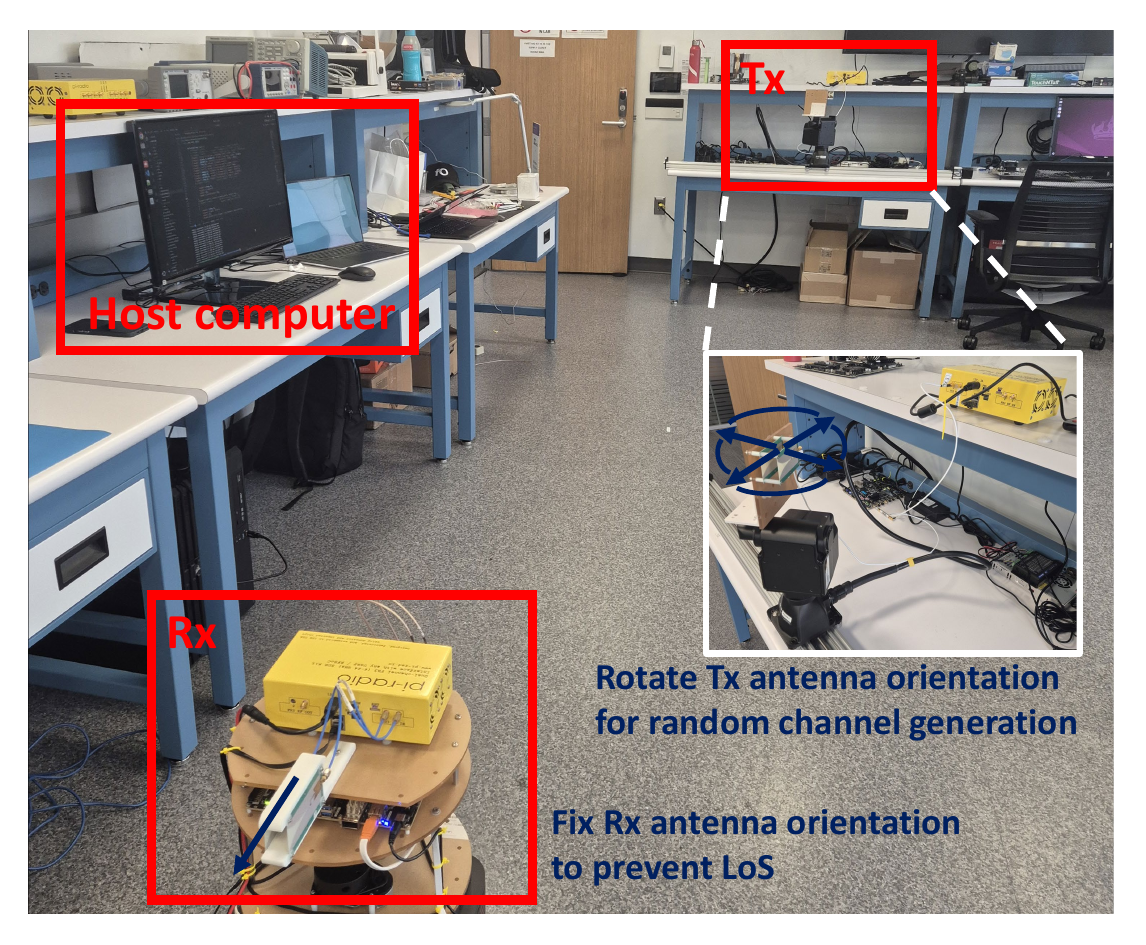}    
\caption{\centering Upper mid-band signal power measurement system.}
\vspace{-1.7em} 
\label{fig:measurement system}
\end{figure}

\section{Upper Mid-Band Measurement System}
\label{sec:measurement}

To empirically validate the multi-band sensing methodologies, the experiments utilize a custom-built $2\times2$ multiple-input multiple-output (MIMO) SDR system engineered specifically for operations within the upper mid-band. The architecture comprises several distinct hardware components working in tandem:

\noindent\textbf{Digital Baseband Processing (RFSoC):} The foundational baseband operations are handled by a Xilinx Zynq UltraScale+ RFSoC (XCZU48DR), which features two digital-to-analog converters (DACs) and four analog-to-digital converters (ADCs). The system functions at a sampling rate of 983.04 MHz and is controlled via a Python-based interface using the PYNQ framework. Within the transmitter's RFSoC, multiple signals are generated digitally and transmitted over the air to be captured by the receiver. These signals are initially outputted at an Intermediate Frequency (IF) centered at 1 GHz. At the receiving end, the captured signals are down-converted back to baseband and transferred to a remote server for post-processing.

\noindent\textbf{Pi-Radio Upper Mid-Band Transceiver Module:} To translate the IF signals to the actual upper mid-band spectrum, the system employs a specialized Pi-Radio transceiver. This module manages the up-conversion at the transmitter and the down-conversion at the receiver between the sub-6 GHz IF and the target radio frequencies. This dual-stage conversion strategy is highly beneficial, as it helps the hardware meet stringent spectral mask and out-of-band emission standards while allowing developers to easily adapt previous sub-6 GHz implementations for higher frequency bands. For this experimental campaign, the RF front end was operated at 10.0 GHz.

\noindent\textbf{Wideband Vivaldi Antennas:} Both the transmitting and receiving nodes are equipped with custom-designed Vivaldi antennas fabricated on printed circuit boards by Pi-Radio. These wideband antennas are designed to support the entire upper mid-band range and are mounted with a physical separation of $2$~cm between them.

\noindent\textbf{Automated Rotational Platform:} To thoroughly test varying angles of departure and arrival, the receiver setup is positioned on an automated rotation platform. A high-precision stepper motor rotates a disk holding the Vivaldi antennas, controlled by an Arduino microcontroller. The Arduino receives orientation commands directly from the main server, streamlining the calibration and measurement process by removing manual adjustments. This mechanical setup achieves an impressive angular precision of roughly $0.1$ degrees.

\section{Results}
\label{Sec:Results}
\subsection{Metrics, configurations, and baselines}

For the simulations, we fix $N=1024$ points and set the target false-alarm probability to $P_{\mathrm{FA}}=10^{-6}$. As the performance metric, we use the intersection-over-union (IoU) error rate, defined as
\begin{align}
e(\mathcal{S},\hat{\mathcal{S}})
= 1 - \frac{|\mathcal{S}\cap \hat{\mathcal{S}}|}{|\mathcal{S}\cup \hat{\mathcal{S}}|},
\end{align}
where $\mathcal{S}$ and $\hat{\mathcal{S}}$ denote the ground-truth and estimated multi-signal interval sets, respectively. 

As baselines, we consider (i) an exhaustive maximum-likelihood (ML) search as an oracle benchmark with $\mathcal{O}(N^2)$ complexity and (ii) learning-based segmentation using U-Net \cite{olaf:2015}, which has been shown to perform well for spectral segmentation and signal detection \cite{nguyen:2024}. For exhaustive ML-based multi-signal interval estimation, we first find the interval with the maximum log-likelihood ratio among the candidates satisfying   $\bar{X}_{\tilde{\mathcal{S}}_k}\ge u_{|\tilde{\mathcal{S}}_k|}$, then remove the selected interval by replacing the corresponding samples with zero power. We repeat these steps until no interval exceeds its threshold, with the maximum number of iterations set to $100$. For the U-Net baselines, we consider two variants: a CNN-based U-Net \cite{olaf:2015} and a U-transformer \cite{Petit:2021}, denoted as `U-Net (CNN)' and `U-Net (Attention)', respectively. The U-Net (CNN) is commonly used for image segmentation because it can extract multi-scale features. The U-Net (Attention) further improves segmentation accuracy by adding multi-head self-attention at the end of the U-Net encoder and multi-head cross-attention at the skip connections, which helps model long-range contextual interactions and spatial dependencies.

\subsection{Simulation data-based verification}

\begin{figure}[t]
\centering
    \begin{subfigure}{0.48\columnwidth}
    \centering
    \includegraphics[width=\columnwidth]{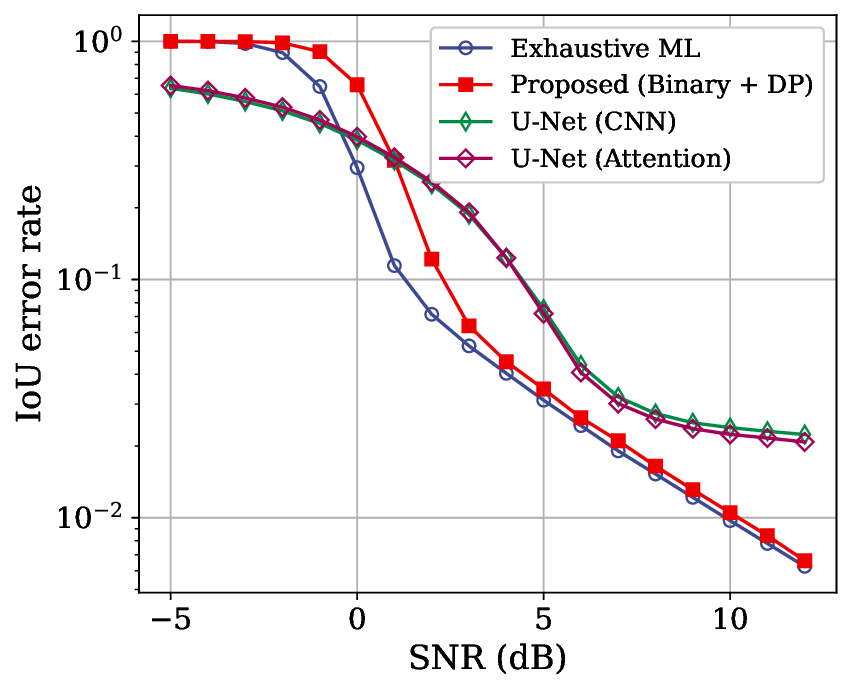}
    \caption{IoU error rate curve with respect to different $\gamma_k \in \{-5,-4,...,12\}$~dB while fixing $|\mathcal{S}_k| = 96$.}
    \label{fig:sim_snr_iou}
    \end{subfigure}
    \begin{subfigure}{0.48\columnwidth}
    \centering
    \includegraphics[width=\columnwidth]{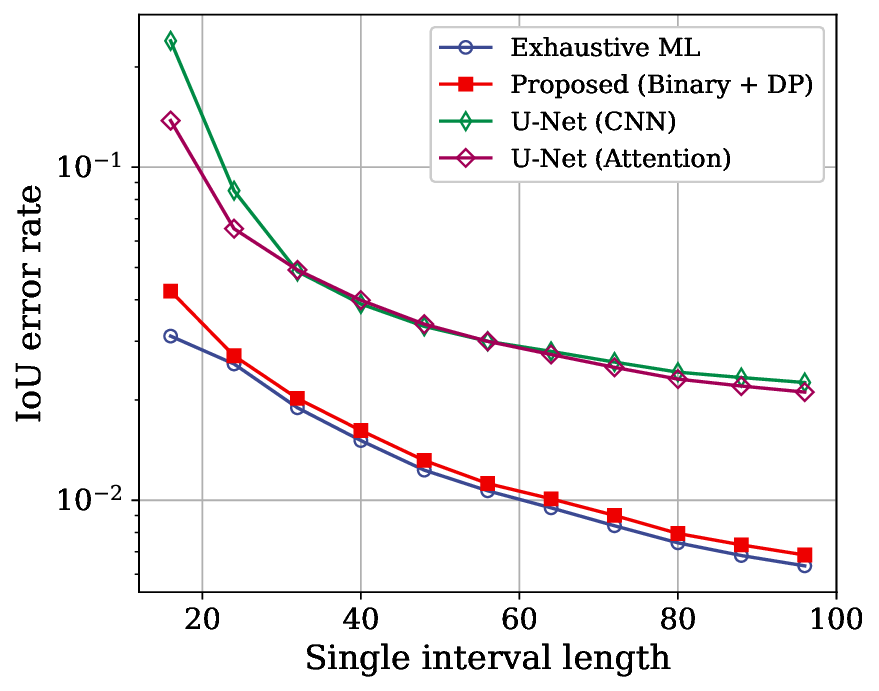}
    \caption{IoU error rate curve with respect to different $|\mathcal{S}_k|\in\{16,24,...,96\}$ while fixing $\gamma = 12$~dB.}
    \label{fig:sim_length_iou}
    \end{subfigure}
\caption{\centering IoU error rate curve with simulation data.}
\vspace{-1.0em}
\label{fig:sim_data}
\end{figure}

To evaluate the IoU error rate under various numbers of signals, bandwidths, and SNRs, we randomly generate $\mathbf{X}$ with a fixed interval length $|\mathcal{S}_k| = 96$ while sweeping the SNR over $\gamma_k \in \{-5,-4,\ldots,12\}$~dB for all $k$ to obtain Fig.~\ref{fig:sim_snr_iou}. We also randomly generate $\mathbf{X}$ with $\gamma_k = 12$~dB while sweeping the interval length over $|\mathcal{S}_k| \in \{16,24,\ldots,96\}$ to obtain Fig.~\ref{fig:sim_length_iou}. These simulations are conducted for each number of signals $K \in \{1,2,3,4\}$, then averaged over $K$ while fixing $G=100$ points. Due to the complexity of exhaustive ML, we assume that the lower and upper bounds of interval length, denoted by $S_{\min}$ and $S_{\max}$, are known and set to $15$ and $129$, respectively.  Accordingly, the complexities of exhaustive ML and the proposed method scale as $\mathcal{O}((S_{\max} - S_{\min})^2)$ and $\mathcal{O}((S_{\max} - S_{\min})\log_2 (S_{\max} - S_{\min}))$, respectively. 
%For U-Net training, we construct $324,000$ training samples by randomly generating $500$ realizations of $\mathbf{X}$ for each combination of $\gamma$, $K$, and $G \in \{20,30,\ldots,100\}$.

Fig.~\ref{fig:sim_snr_iou} shows the dependence of the IoU error rate on SNR. In addition, Table~\ref{table:elapsed time} summarizes the elapsed time and ratio with respect to exhaustive ML, measured on an Intel(R) Xeon(R) Gold 6146 CPU for various $K \in \{1,2,3,4\}$, each averaged on $|\mathcal{S}_k| \in \{16,32,48,64\}$. As expected, the IoU error rates of all methods decrease as the SNR increases, and exhaustive ML achieves the lowest IoU error rate at high SNRs. The proposed method closely tracks exhaustive ML, with only a small performance gap, while requiring only $0.74\%$ of exhaustive ML's elapsed time on average. This demonstrates that the proposed method achieves near-optimal performance at substantially lower computational cost. For the U-Net-based baselines, the U-Net (Attention) slightly outperforms the U-Net (CNN) because an arbitrary number of signals makes the segmentation problem require more global contextual reasoning, for which attention structures may be better suited than a U-Net (CNN), which primarily captures local information. However, both U-Net variants yield noticeably higher IoU error rates than the proposed method at high SNRs, and require longer runtimes. Next, Fig.~\ref{fig:sim_length_iou} shows the dependence of the IoU error rate on $|\mathcal{S}_k|$. As $|\mathcal{S}_k|$ increases, the IoU error rate decreases for all methods, while the performance ordering observed in Fig.~\ref{fig:sim_snr_iou} at high SNR is maintained. This is because signal power fluctuations can confuse interval decisions, especially when the signal intervals are short.

\begin{table}[t]
    \centering
    \small
    \captionsetup{justification=centering, labelsep=newline, font={smaller,sc}}
    \caption{Averaged elapsed time (Unit: Sec) and ratio with respect to exhaustive ML.}
    \label{table:elapsed time}
    \resizebox{\columnwidth}{!}{
    \begin{tabular}{c|c|c|c|c}
    \toprule
    $K$ & Exhaustive ML &  \textbf{Proposed (Binary + DP)} & U-Net (CNN) & U-Net (Attention) \\\hline
    
    $1$ & $0.2096$ $(100\%)$ & $\mathbf{0.0016}$ $\mathbf{(0.76\%)}$ & $0.0062$ $(2.96\%)$ & $0.0187$ $(8.92\%)$ \\\hline
    
    $2$  & $0.2931$ $(100\%)$ & $\mathbf{0.0023}$ $\mathbf{(0.78\%)}$ & $0.0064$ $(2.18\%)$ & $0.0189$ $(6.45\%)$ \\\hline
    
    $3$ &  $0.3821$ $(100\%)$ & $\mathbf{0.0028}$ $\mathbf{(0.73\%)}$ & $0.0064$ $(1.67\%)$ & $0.0190$ $(4.97\%)$ \\\hline
    
    $4$ & $0.4716$ $(100\%)$  & $\mathbf{0.0033}$ $\mathbf{(0.70\%)}$ & $0.0062$ $(1.31\%)$ & $0.0186$ $(3.94\%)$ \\\hline
    \end{tabular}}
    \vspace{-1.3em}
\end{table}

\subsection{Experimental data-based verification}

\begin{figure}[t]
\centering
    \begin{subfigure}{0.5\columnwidth}
    \centering
    \includegraphics[width=\columnwidth]{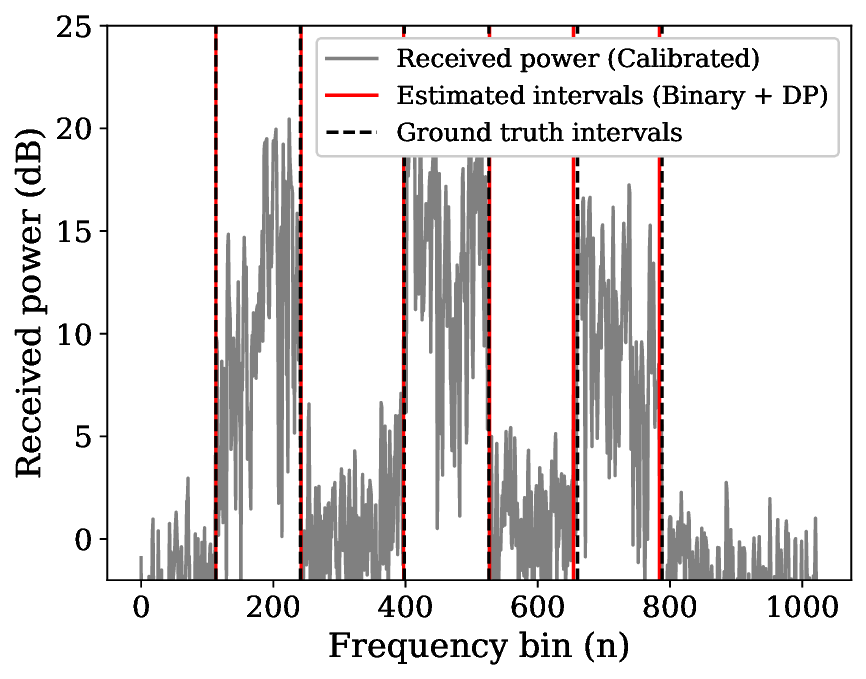}
    \caption{Experimental data example with ground truth and estimated intervals.}
    \label{fig:measurement_example}
    \end{subfigure}
    \vspace{-0.5em}
    \begin{subfigure}{0.48\columnwidth}
    \centering
    \includegraphics[width=\columnwidth]{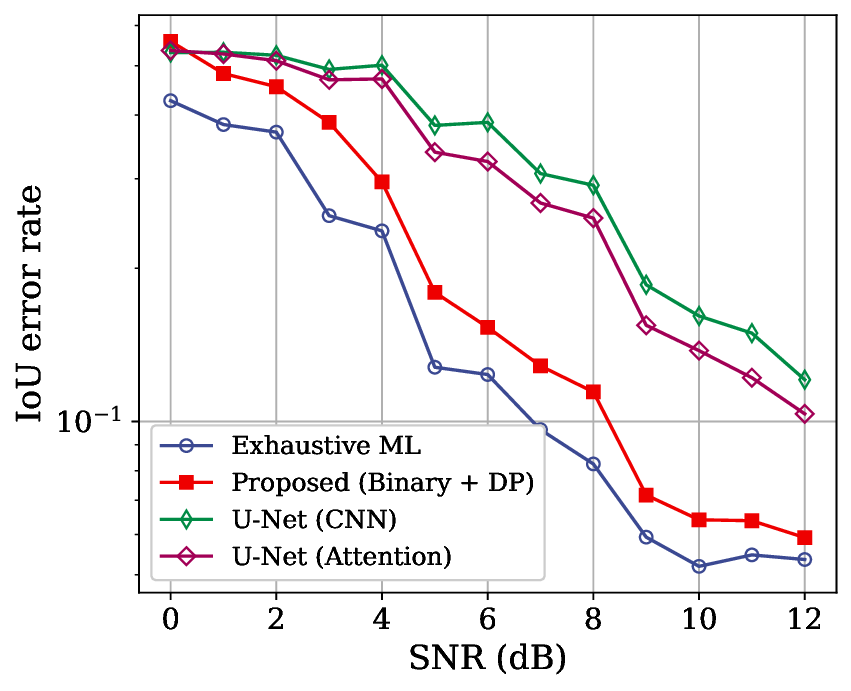}
    \caption{IoU error rate curve with respect to different $\gamma_k \in \{0,1,...,12\}$~dB.}
    \label{fig:sim_snr_iou_measurement}
    \end{subfigure}
    \vspace{-0.8em}
\caption{\centering Experimental data example and IoU error rate curve.}
\vspace{-1.0em}
\label{fig:meas_data}
\end{figure}

We further evaluate the proposed algorithm using experimental data obtained from the Pi-Radio SDR introduced in Sec.~\ref{sec:measurement}, again having $N = 1024$ points while setting the bandwidth of adjacent bins as $960$~kHz. For the U-Net baselines, we use the same models as in Fig.~\ref{fig:sim_data}, which were trained using simulation data. For calibration of the measurement data, we first estimate the SNR as $\sum_{n \in \mathcal{S}} X_n / |\mathcal{S}| - \sum_{n \notin \mathcal{S}} X_n / |\mathcal{S}^c|$, where $\mathcal{S}^c$ is the complement of $\mathcal{S}$, assuming that $\{\gamma_k\}_{k=1}^K$ are identical. We then scale the measurements separately inside and outside the ground-truth signal interval, i.e., for $n \in \mathcal{S}$ and $n \notin \mathcal{S}$, such that their empirical averages satisfy the expectation conditions in \eqref{eq:expectation cases}. An example of the calibrated measurement data is presented in Fig.~\ref{fig:measurement_example}, together with the ground-truth and estimated signal intervals. We then average the IoU error rates over samples whose estimated SNR lies in $\gamma_k \in [\gamma,\gamma+1)$, $K \in \{1,2,...,5\}$, and $|\mathcal{S}_k| \in \{16,64,128\}$, then report the result at $\gamma_k \in \{0,1,2,\ldots,12\}$~dB as illustrated in Fig.~\ref{fig:sim_snr_iou_measurement}.

Compared with the simulation results in Fig.~\ref{fig:sim_snr_iou}, the IoU error rates increase for all methods because the measurement data does not perfectly follow the theoretical model. Nevertheless, the relative performance trend remains the same: exhaustive ML performs best, followed by the proposed method, U-Net (Attention), and U-Net (CNN). The U-Net baselines experience larger performance degradation because the measurement data is out-of-distribution relative to the simulation data used for training. Therefore, we conclude that the proposed method preserves its near-ML detection capability even under practical measurement-model mismatch. Moreover, compared with the U-Net-based baselines, the proposed method exhibits stronger robustness to out-of-distribution measurement data.

\section{Conclusion}
\label{Sec:Conclusion}
In this paper, we introduced a computationally efficient binary search method to estimate the intervals of the signals when the number, bandwidths, and SNRs of the signals are unknown. To reduce the complexity, we adopt a strategy to first find the dyadic intervals through GLRT, find the optimal set of dyadic intervals that maximizes the sum of the log-likelihood through DP, and refine each interval, whose overall complexity is much smaller than exhaustive ML, resulting in only $0.74\%$ of the elapsed time under our simulation setup. We verified the algorithm through both simulation and experimental data. We used an SDR for the upper mid-band experiment developed by Pi-Radio, verifying the applicability of the proposed algorithm in a real environment by showing its robustness under out-of-distribution data. Future work will include extension to multiple dimensions and exploiting frequency selectivity for more accurate SS while maintaining low computational complexity.

%\newpage
% \appendix
% \appendices

\bibliographystyle{refs/IEEEtran}
\bibliography{refs/IEEEfull,refs/bibliography}{}

%\newpage
%\onecolumn

\end{document}